\documentclass[useAMS,usenatbib,letters]{mn2e}
\usepackage{epsf}

\title[How many arcminute-separation lenses are expected in the 2dF QSO
survey?] 
{How many arcminute-separation lenses are expected in the 2dF QSO survey?}
\author[M. Oguri]
{Masamune Oguri$^{1}$\thanks{E-mail:
oguri@utap.phys.s.u-tokyo.ac.jp}\\
$^{1}$Department of Physics, School of Science, University of Tokyo,
Tokyo 113-0033, Japan}

\voffset- .65in

\begin{document}

\date{\today}

\pagerange{L\pageref{firstpage}--L\pageref{lastpage}} \pubyear{2002}

\maketitle

\label{firstpage}

\begin{abstract}
Wide separation lensing statistics offer information about the
density profile and abundance of dark halos. Recently a possible
discovery of six quasar pairs, which may be lensed multiple images, was
reported by Miller et al. (2003). These pairs are selected from a
catalog of the 2dF quasar (QSO) survey comprising 22163 quasars. We
calculate expected numbers of lensed quasars taking account of 
the redshift and magnitude distributions of the quasar catalog. Given
some of the six pairs are genuine lensed systems, we put interesting
constraints on the inner slope of dark halos, $\Omega_0$, and
 $\sigma_8$. We show that the detection of even one lens with separation
$>30''$ is marginally consistent with models that have cuspy inner
density profile, $\rho\propto r^{-1.5}$, and very large $\sigma_8$,
$\sigma_8 \ga 1.2$ for $\Omega_0=0.3$. To reconcile with constraints
from X-ray clusters or cosmic shear, much lower $\Omega_0$ and much
higher $\sigma_8$ are needed, although such high $\sigma_8$ seems too
extreme. The shallower inner density profile  $\rho\propto r^{-1}$ is
hardly acceptable. In particular, the expected number of lenses with
separation $>200''$ is too small to explain the discovery of such
anomalously wide separation lens systems. These results imply that
we miss some important systematic effects, there is a problem in the cold
dark matter scenario, or none of these six quasar pairs is likely to be
lensed images. 
\end{abstract}


%%%%%%%%%%%%%%%%%%%%%%%%%%%%%%%%%%%%%%%%%%%%%%%%%%%%%%%%%%%
\section{Introduction}\label{sec:intro}
%%%%%%%%%%%%%%%%%%%%%%%%%%%%%%%%%%%%%%%%%%%%%%%%%%%%%%%%%%%
The Cold Dark Matter (CDM) scenario predicts the existence of cuspy dark
halos, and thus is expected to produce significant numbers of wide
separation lenses ($\theta\ga 6''$). Statistics of such wide separation
lenses are known to be a powerful tool to probe the abundance
\citep{narayan88,kochanek95} and the density profile of dark halos
\citep*{maoz97,wyithe01,keeton01,sarbu01,takahashi01,li02,oguri02b}. Although a
number of radio surveys has tried, they could not find wide separation
lensed quasars \citep*[e.g.,][]{phillips01a,phillips01b,ofek01,ofek02}.
The lack of wide separation lenses, however, does not conflict with the
CDM scenario because the expected lensing rate is significantly smaller
than that of small separation lensing ($\theta\sim 1''$). 

Recently, \citet{miller03} reported that they found six quasar pairs
which may be lensed multiple images in the 2dF quasar (QSO) catalog
comprising 22163 quasars, although the number of lensed quasars may be
even larger than this because they have done follow-up for only 11 quasar
pairs among 38 quasars selected as possible lens candidates. These
systems were identified to be lensed images from the detailed comparison
of quasar spectra. The separations of all these quasar pairs are larger
than $30''$, and the separations of some pairs reach even $\sim 200''$.
If this surprising result is true, it offers a lot of information about
dark halos. In this {\it Letter}, we calculate the expected number of
arcminute-separation lenses ($\theta>30''$) in the 2dF QSO survey. We
use the realistic density  profile predicted in the CDM scenario, and
take account of the redshift and magnitude distribution of the quasar
catalog. Therefore our results can be directly compared with the
observation. We show that the existence of such wide separation lenses
in the 2dF QSO catalog is marginally consistent with the ``concordance''
cosmology if the value of only $\sigma_8$ is very large,
$\sigma_8\ga1.2$ for $\Omega_0=0.3$. This means that we can strongly
constrain dark halo properties if some of quasar pairs reported by
\citet{miller03} are truly gravitational lens systems. Throughout the
paper, we assume a flat universe  $\Omega_0+\lambda_0=1$ and
$h=H_0/(100{\rm km\,s^{-1}Mpc^{-1}})=0.7$. 
 
%%%%%%%%%%%%%%%%%%%%%%%%%%%%%%%%%%%%%%%%%%%%%%%%%%%%%%%%%%%
\section{Calculation of Probability Distributions}
%%%%%%%%%%%%%%%%%%%%%%%%%%%%%%%%%%%%%%%%%%%%%%%%%%%%%%%%%%%
The lensing probability distribution at wide separation reflects the
properties of dark halos, rather than galaxies
\citep{nakamura97,keeton98,oguri02b}. The halo density profiles
predicted by recent N-body simulations may be parameterized as a
one-parameter family, the generalized NFW profile \citep{zhao96,jing00a}:  
%%%%%%%%%%%%%%%%%%%%%%%%%%%%%%%%%%%%%%%%%%%%%%%%%%%%%%%%%%%%%%%%%%%%%%%%
\begin{equation}
 \rho(r)=\frac{\rho_{\rm crit}(z)\delta_{\rm c}(z)}
{\left(r/r_{\rm s}\right)^\alpha\left(1+r/r_{\rm s}\right)^{3-\alpha}},
\label{nfw}
\end{equation}
%%%%%%%%%%%%%%%%%%%%%%%%%%%%%%%%%%%%%%%%%%%%%%%%%%%%%%%%%%%%%%%%%%%%%%%%
where $\rho_{\rm crit}(z)$ is a critical density at $z$. While the
correct value of $\alpha$ is still unclear, the existence of a cusp with
$1\la\alpha\la 1.5$ has been established in recent N-body simulations
\citep*{navarro96,moore99,jing00a,fukushige01}. For definiteness, in
this paper we consider two cases, $\alpha=1$ and $\alpha=1.5$, which
cover the range of the CDM predictions. The scale radius $r_{\rm s}$ is
related to the concentration parameter $c_{\rm vir}(M, z)\equiv r_{\rm
vir}(M, z)/r_{\rm s}(M, z)$. Then the characteristic density $\delta_{\rm
c}(z)$ is given in terms of the concentration parameter \citep*[see
e.g.,][]{oguri01}. The lens equation of the generalized NFW profile has
three solutions if $|\bmath{\eta}|<\eta_{\rm r}$, where $\bmath{\eta}$
is the position of a source in the source plane and $\eta_{\rm r}$ is a
radius of the radial caustic in the source plane. The image separation
is defined between the outer two solutions and is approximated as
$\theta(M, z_{\rm S}, z_{\rm L})\simeq 2\xi_{\rm t}/D_{\rm OL}$, where
$\xi_{\rm t}$ is a radius of the tangential critical curve in the lens
plane, $D_{\rm OL}$ denotes the angular diameter distance from the
observer to the lens plane, and $z_{\rm S}$ and $z_{\rm L}$ indicate
the redshifts of the source and lens, respectively.

The concentration parameter $c_{\rm vir}$ is one of the most important
parameter in the generalized NFW density profile (eq. [\ref{nfw}]). 
In numerical simulations, it has been found that the concentration
parameter depends on the mass $M$ and redshift $z$ of halos
\citep[e.g.,][]{bullock01}. Moreover, it shows considerable scatter
which reflects the difference in the formation epoch \citep{wechsler02}.
The scatter in the concentration parameter is well described by a
log-normal distribution. For the median of concentration parameter
$c_{\rm vir,med}$, we adopt the mass and redshift dependence reported by
\citet{bullock01}: 
%%%%%%%%%%%%%%%%%%%%%%%%%%%%%%%%%%%%%%%%%%%%%%%%%%%%%%%%%%%%%%%
\begin{equation}
 c_{\rm vir,med}(M, z)=(2-\alpha)\frac{8}{1+z}\left(\frac{M}{10^{14}h^{-1}M_{\odot}}\right)^{-0.13}.
\label{conmed}
\end{equation}
%%%%%%%%%%%%%%%%%%%%%%%%%%%%%%%%%%%%%%%%%%%%%%%%%%%%%%%%%%%%%%%
Note that this fitting form is slightly different from the one
\citet{bullock01} originally proposed, and is correct at $M\sim
10^{14}h^{-1}M_{\odot}$. 
A factor $2-\alpha$ in equation (\ref{conmed}) gives a natural way
to generalize $\alpha\neq 1$ \citep{keeton01,jing02}. The scatter of the
concentration parameter $\sigma_c$ is also important element in
gravitational lens statistics \citep{keeton01}. We adopt the value 
$\sigma_c=0.3$ which has been obtained from N-body simulations
\citep{jing00b,bullock01,wechsler02,jing02}. 

The probability that a source at redshift $z_{\rm S}$ and having the
absolute luminosity $L$ is observed as multiply lensed system with
separation larger than $\theta$ is given by
%%%%%%%%%%%%%%%%%%%%%%%%%%%%%%%%%%%%%%%%%%%%%%%%%%%%%%%%%%%%%%%%%%%%%%%%
\begin{eqnarray}
P^{\rm B}(>\!\theta; z_{\rm S}, L)&=&\int_0^{z_{\rm S}}dz_{\rm L}\int_{M_{\rm min}}^\infty dM \nonumber\\
&&\times\sigma_{\rm lens}B\frac{c\,dt}{dz_{\rm L}}(1+z_{\rm L})^3\frac{dn}{dM},
\label{cpd_bias}
\end{eqnarray}
%%%%%%%%%%%%%%%%%%%%%%%%%%%%%%%%%%%%%%%%%%%%%%%%%%%%%%%%%%%%%%%%%%%%%%%%
where $B$ is the magnification bias \citep{turner80}:
%%%%%%%%%%%%%%%%%%%%%%%%%%%%%%%%%%%%%%%%%%%%%%%%%%%%%%%%%%%%%%%%%%%%%%%%
\begin{equation}
B
=\frac{2}{y_{\rm r}^2\Phi(z_{\rm S}, L)}\int_0^{y_{\rm r}}dy\,y\,\Phi(z_{\rm S}, L/\mu(y))\frac{1}{\mu(y)},
\label{biasfactor}
\end{equation}
%%%%%%%%%%%%%%%%%%%%%%%%%%%%%%%%%%%%%%%%%%%%%%%%%%%%%%%%%%%%%%%%%%%%%%%%
with $\Phi(z_{\rm S}, L)$ being the luminosity function of sources. The
lensing cross section $\sigma_{\rm lens}$ is simply given by the area
encompassed by the radial caustic, $\sigma_{\rm lens}=\pi\eta_{\rm
r}^2D_{\rm OL}^2/D_{\rm OS}^2$, where $D_{\rm OS}$ indicates the angular
diameter distance from the observer to the source plane. The lower limit
of mass integral $M_{\rm min}$ is related to $\theta$ as
$\theta=\theta(M_{\rm min}, z_{\rm S}, z_{\rm L})$. The magnification
bias (eq. [\ref{biasfactor}]) should be calculated for the faintest of 
the two images, because both lensed images must appear above the flux
limit of the 2QZ survey \citep{miller03}.  We use an approximation
of the magnification factor $\mu(y)$ which was derived by \citet{oguri02a}. 

Since wide separation lenses with $\theta>30''$ are considered to be 
generated by massive clusters, we should choose the mass function of
dark halos ($dn/dM$ in eq. [\ref{cpd_bias}]) carefully. We adopt
equation (B3) of \citet{jenkins01} which well agrees with simulated
high-mass halo abundance \citep{evrard02,komatsu02,hu03,pierpaoli03}.
Note that $\rho=180\Omega(z)\rho_{\rm crit}(z)$ should be used as the
mean overdensity when one adopts this mass function. 
%%%%%%%%%%%%%%%%%%%%%%%%%%%%%%%%%%%%%%%%%%%%%%%%%%%%%%%%%%%
\section{Quasar Catalog}
%%%%%%%%%%%%%%%%%%%%%%%%%%%%%%%%%%%%%%%%%%%%%%%%%%%%%%%%%%%
To make a precise prediction which can be directly compared with
observed lensing rate, we must properly take account of the redshift and
magnitude distributions. Since the whole sample used to search lensed
quasars is not publicly available, we instead use the 2dF 10k catalog
comprising 
$\sim 10000$ quasars \citep{croom01} which is a part of the whole
sample. Then predicted numbers of wide separation lensing are calculated
as follows. First, from the catalog we extract the numbers of quasars
$N(z_i,b_j)$ which are located $z_i-\Delta z/2<z_i<z_i+\Delta z/2$ and
have magnitude $b_j-\Delta b/2<b_j<b_j+\Delta b/2$. We use $\Delta
z=0.1$ and $\Delta b=0.2$. The average probability that quasars are
lensed with separations larger than $\theta$ is then given by  
%%%%%%%%%%%%%%%%%%%%%%%%%%%%%%%%%%%%%%%%%%%%%%%%%%%%%%%%%%%%%%%%%%%%%%%%
\begin{equation}
P_{\rm lens}(>\!\theta)=\frac{\sum_{z_i}\sum_{b_j}N(z_i,b_j)P(>\!\theta; z_i, L(b_j))}{\sum_{z_i}\sum_{b_j}N(z_i,b_j)},
\end{equation}
%%%%%%%%%%%%%%%%%%%%%%%%%%%%%%%%%%%%%%%%%%%%%%%%%%%%%%%%%%%%%%%%%%%%%%%%
where $L(b)$ is the $B$-band absolute luminosity corresponds to $b$.
When the quasar continuum spectrum is described by a power law,
$f_\nu\propto\nu^{-\alpha_{\rm s}}$, the K-correction can be
approximated as $K(z)=-2.5(1-\alpha_{\rm s})\log(1+z)$. We assume
$\alpha_{\rm s}=0.5$ to calculate the K-correction. The total number of
lensed quasars expected in the 2dF QSO survey is 
%%%%%%%%%%%%%%%%%%%%%%%%%%%%%%%%%%%%%%%%%%%%%%%%%%%%%%%%%%%%%%%%%%%%%%%%
\begin{equation}
N_{\rm lens}(>\!\theta)=N_{\rm QSO}P_{\rm lens}(>\!\theta),
\label{num}
\end{equation}
%%%%%%%%%%%%%%%%%%%%%%%%%%%%%%%%%%%%%%%%%%%%%%%%%%%%%%%%%%%%%%%%%%%%%%%%
where $N_{\rm QSO}=22163$ is the total number of quasars in the whole
sample. 

The luminosity function of quasars is needed to compute magnification
bias. We adopt the double power law luminosity function:
%%%%%%%%%%%%%%%%%%%%%%%%%%%%%%%%%%%%%%%%%%%%%%%%%%%%%%%%%%%%%%%%%%%%%%%%
\begin{equation}
\Phi(z,L)dL=\frac{\Phi_*}{[L/L_*(z)]^{\beta_h}+[L/L_*(z)]^{\beta_l}}\frac{dL}{L_*(z)}.
\end{equation}
%%%%%%%%%%%%%%%%%%%%%%%%%%%%%%%%%%%%%%%%%%%%%%%%%%%%%%%%%%%%%%%%%%%%%%%%
We assume pure luminosity evolution models with $L_*(z)\propto
10^{k_1z+k_2z^2}$, and use a best-fitting model which was derived by
\citet{boyle00} in the $(\Omega_0,\lambda_0)=(0.3,0.7)$ universe:
$\beta_h=3.41$, $\beta_l=1.58$, $k_1=1.36$, $k_2=-0.27$, and
$M_*=-21.15+5\log h$. 

%%%%%%%%%%%%%%%%%%%%%%%%%%%%%%%%%%%%%%%%%%%%%%%%%%%%%%%%%%
%%%%%%%%%%%%%%%%%%%%%%%%%%%%%%%%%%%%%%%%%%%%%%%%%%%%%%%%%%%
\section{Results}
%%%%%%%%%%%%%%%%%%%%%%%%%%%%%%%%%%%%%%%%%%%%%%%%%%%%%%%%%%%
%%%%%%%%%%%%%%%%%%%%%%%%%%%%%%%%%%%%%%%%%%%%%%%%%%%%%%%%%%%

%%%%%%%%%%%%%%%%%%%%%%%%%%%%%%%%%%%%%%%%%%%%%%%%%%%%%%%%%%
\subsection{Numbers of Lenses}
%%%%%%%%%%%%%%%%%%%%%%%%%%%%%%%%%%%%%%%%%%%%%%%%%%%%%%%%%%

%%%%%%%%%%%%%%%%%%%%%%%%%%%%%%%%%%%%%%%%%%%%%%%%%%%%%%%%%%%%%%%%%%%%%
\begin{figure}
  \begin{center}
    \epsfxsize=8.1cm
    \epsfbox{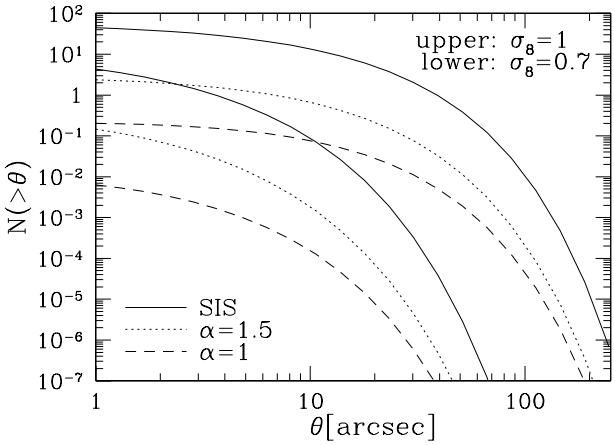}
  \end{center}
    \caption{The predicted number distribution of image separations in
 the 2dF QSO survey (eq. [\ref{num}]). For the density profile of lens
 objects, both $\alpha=1.5$ and $\alpha=1$ are considered (see eq.
 [\ref{nfw}]). The numbers assuming Singular Isothermal Sphere (SIS)
 density profile, $\rho\propto r^{-2}$, are also plotted for reference.
 The cosmological models adopted in these plots are
 $(\Omega_0,\sigma_8)=(0.3,0.7)$ and $(\Omega_0,\sigma_8)=(0.3,1.0)$ in
 the flat universe ($\Omega_0+\lambda_0=1$). }
\label{fig:sep}
\end{figure}
%%%%%%%%%%%%%%%%%%%%%%%%%%%%%%%%%%%%%%%%%%%%%%%%%%%%%%%%%%%%%%%%%%%%%

First we plot the predicted number distribution of image separation in
the 2dF QSO survey in Figure \ref{fig:sep}. The cases that the density
profile of lens objects is described by the Singular Isothermal Sphere
(SIS) are also shown for reference. The value of $\sigma_8$ has been
constrained from X-ray clusters or cosmic shear, but resultant values
show discrepancies among papers, ranging from $\sim 0.7$ to $\sim 1.0$
for $\Omega_0\simeq 0.3$ \citep[e.g.,][]{pierpaoli03}. Therefore we plot 
both $\sigma_8=0.7$ and $\sigma_8=1$ models. Figure \ref{fig:sep}
clearly indicates that the predicted numbers of lenses strongly depend
on both density profile ($\alpha$) and the abundance ($\sigma_8$) of
dark halos. In particular, numbers of arcminute-separation lensed
quasars are highly sensitive to $\sigma_8$ because such very wide
separation lenses are mainly produced by massive clusters. We find that
expected numbers of lensed quasars with image separation larger than
$200''$ is quite small, $N_{\rm lens}(>200'')<10^{-6}$, while 4 quasar
pairs which are likely to be lensed were found \citep{miller03}. We
further find that even an ``extreme'' model with $\alpha=1.5$ and
$\sigma_8=1.4$ can produce only $\sim 3\times10^{-4}$ lenses with
$\theta>200''$ on average. On the other hand, less wide separation
lenses ($\theta\sim 30''$) may be statistically possible if both
$\alpha$ and $\sigma_8$ are large. 

%%%%%%%%%%%%%%%%%%%%%%%%%%%%%%%%%%%%%%%%%%%%%%%%%%%%%%%%%%%%%%%%%%%%%
\begin{figure}
  \begin{center}
    \epsfxsize=8.1cm
    \epsfbox{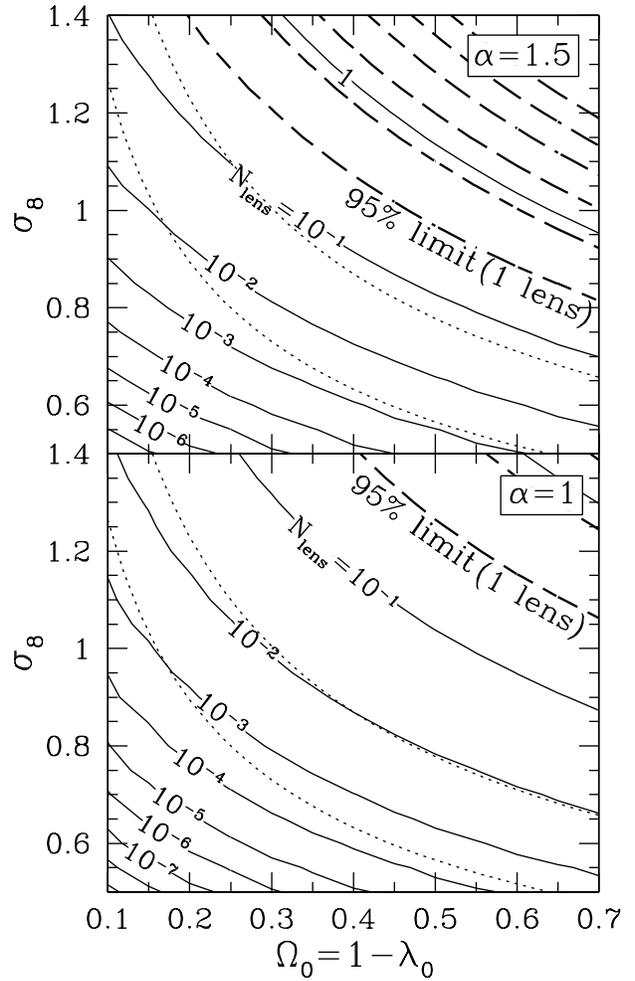}
  \end{center}
    \caption{Contours of $N_{\rm lens}(>30'')$ in the
 $\Omega_0$-$\sigma_8$ plane. Constraints (95\% confidence level) from
 the detection of wide separation lenses are shown by dashed lines. From
 lower to upper of dashed lines, the number of genuine lens systems is
 assumed to be 1, 2, 3, 4, 5, 6. Dotted lines indicate the recent
 constraints on $\Omega_0$ and $\sigma_8$ from X-ray clusters or cosmic
 shear which can be approximately expressed as
 $0.4<\sigma_8\Omega_0^{0.5}<0.55$.}   
\label{fig:os}
\end{figure}
%%%%%%%%%%%%%%%%%%%%%%%%%%%%%%%%%%%%%%%%%%%%%%%%%%%%%%%%%%%%%%%%%%%%%

To see what constraint can be put from the detection of wide separation
lenses, in Figure \ref{fig:os} we plot contours of $N_{\rm lens}(>30'')$
in the $\Omega_0$-$\sigma_8$ plane. We also plot constraints from the
observed number of lenses assuming that some of six pairs reported by
\citet{miller03} are true lens systems. These constraints are calculated 
from the Poisson distribution $P(k|N)=N^ke^{-N}/k!$, where $k$ is
the observed number and $N$ is the expectation. Given the numbers of
true lenses range from $k=1$ to $6$, the lower limits of $N$ (95\%
confidence level) become $0.35$, $0.82$, $1.36$, $1.97$, $2.61$, and
$3.29$, respectively. This figure indicates that the detection of even
one lens with $\theta>30''$ in the 2dF QSO catalog needs very high
$\sigma_8$, $\sigma_8\ga 1.2$ for $\Omega_0=0.3$ and $\alpha=1.5$.
Our constraint and constraints from X-ray clusters or cosmic shear which
are approximated as $0.4<\sigma_8\Omega_0^{0.5}<0.55$ \citep[see
e.g.,][]{pierpaoli03} shows marginal agreement if $\Omega_0\la 0.15$,
$\sigma_8\ga1.5$, and $\alpha=1.5$, although this solution seems too
extreme.  In particular, models with $\alpha=1.0$ are hardly acceptable
because they need unusually high $\sigma_8$ or $\Omega_0$ in order to
produce lenses with $\theta>30''$. However, this result is somewhat
embarrassing because the lack of wide separation lensing in other
surveys \citep*[e.g.,][]{phillips01a,phillips01b,ofek01,ofek02} already 
puts the upper limit of the lensing rate. For instance, our model with
$\alpha=1.5$, $\Omega_0=0.3$, and $\sigma_8=1.2$ predicts a lensing rate
$P(>6'')\sim 3\times 10^{-4}$ at $z\sim 1.3$ which is marginally
consistent with the upper limit of the lensing rate in
\citet{phillips01b}, $P(>6'')\la 3\times 10^{-4}$ at 95\% confidence limit. 

Note that these constraints are for the case that only one of six quasar
pairs is a genuine lens system; if the number of lenses is more than
one, the situation becomes worse. In this case, the discrepancy between
strong lensing constraints and X-ray/shear constraints becomes more
serious. Moreover, this result may conflict with other surveys which
could not detect wide separation lenses.  

%%%%%%%%%%%%%%%%%%%%%%%%%%%%%%%%%%%%%%%%%%%%%%%%%%%%%%%%%%
\subsection{Theoretical Uncertainties}
%%%%%%%%%%%%%%%%%%%%%%%%%%%%%%%%%%%%%%%%%%%%%%%%%%%%%%%%%%

\begin{table}
 \caption{Theoretical uncertainties in the expected numbers of lensed
 quasars. The fiducial model has $\alpha=1.5$ and
 $(\Omega_0,\sigma_8)=(0.3,1.0)$. ``Press \& Schechter MF'' and ``Sheth
 \& Tormen MF'' models are same as the fiducial model except for using
 the mass function derived by \citet{press74} and \citet{sheth99}
 instead of that fitted by \citet{jenkins01}, respectively. In ``Croom
 et al. LF'' model, we use the quasar luminosity function from the 10k
 catalog \citep{croom01}. Note that this luminosity function has fairly
 shallower bright- and faint-end slopes, $\beta_h=3.28$ and
 $\beta_l=1.08$. \label{table:unc}}
 \begin{tabular}{@{}ccc}
  \hline
  Model & $N_{\rm lens}(>30'')$ &  $N_{\rm lens}(>200'')$\\
  \hline
  Fiducial Model & $7.9\times 10^{-2}$ & $1.8\times10^{-7}$\\
  Press \& Schechter MF & $1.2\times 10^{-1}$ & $1.6\times10^{-7}$\\
  Sheth \& Tormen MF & $2.8\times 10^{-1}$ & $8.3\times10^{-6}$\\
  Croom et al. LF & $4.4\times 10^{-2}$ & $1.1\times10^{-7}$\\
  \hline
 \end{tabular}
\end{table}

We also examine possible theoretical uncertainties except for the
uncertainties of the density profile and cosmological parameters. More
specifically, we examine the uncertainties of the halo mass function and
the quasar luminosity function. The result is summarized in Table
\ref{table:unc}. First we adopt the mass functions of \citet{press74}
and \citet{sheth99} and see how the number of lensed quasars changes.
We find that the uncertainty of the mass function is fairly large. In
particular, the mass function of \citet{sheth99} predicts more than
three times as large number of lenses with $\theta>30''$ as our fiducial
model. This is because the mass function of \citet{sheth99} seems to
overestimate the number density of massive halos \citep{jenkins01,hu03}.
Moreover, we point out that the number of anomalously wide separation
lenses ($\theta>200''$) is much more sensitive to the choice of the mass
function. Therefore, in the statistics of such anomalously wide separation
lenses the uncertainty of the mass function should be carefully examined.

Since the magnification bias is sensitive to the slope of adopted quasar
luminosity function, next we examine the uncertainty of the quasar
luminosity function using the luminosity function from the 10k catalog
\citep{croom01}. This luminosity function has somewhat shallower slopes,
$\beta_h=3.28$ and $\beta_l=1.08$, compared with our fiducial model.
We find that this uncertainty is less than factor 2 and is not so large as
to change our main results, because $\alpha$, $\Omega_0$, and $\sigma_8$
can change the number of wide separation lenses by orders of magnitude. 

%%%%%%%%%%%%%%%%%%%%%%%%%%%%%%%%%%%%%%%%%%%%%%%%%%%%%%%%%%
\subsection{Expected Time Delays between Images}
%%%%%%%%%%%%%%%%%%%%%%%%%%%%%%%%%%%%%%%%%%%%%%%%%%%%%%%%%%

%%%%%%%%%%%%%%%%%%%%%%%%%%%%%%%%%%%%%%%%%%%%%%%%%%%%%%%%%%%%%%%%%%%%%
\begin{figure}
  \begin{center}
    \epsfxsize=8.1cm
    \epsfbox{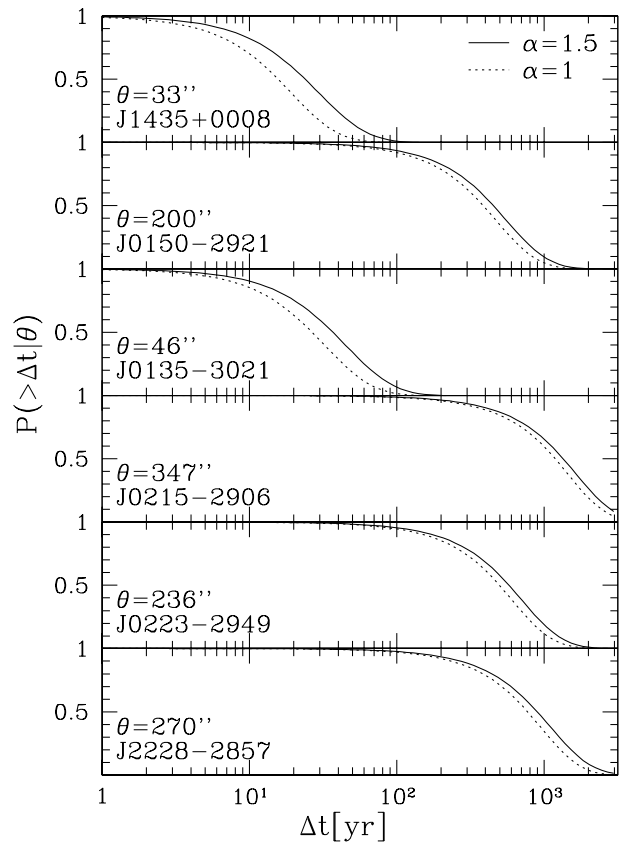}
  \end{center}
    \caption{The conditional probability distributions of possible time
 delays are shown for each quasar pair. The definition and calculation of 
 $P(>\Delta t|\theta)$ was shown by \citet{oguri02a}. The redshift and
 magnitude for each image are taken into account. For the
 cosmological model, we adopt $(\Omega_0,\sigma_8)=(0.3,1.0)$. }
\label{fig:tdelay}
\end{figure}
%%%%%%%%%%%%%%%%%%%%%%%%%%%%%%%%%%%%%%%%%%%%%%%%%%%%%%%%%%%%%%%%%%%%%

The main drawback of wide separation lensing statistics is that it is
hard to recognize quasar pairs as gravitational lens systems due to
large time delays between images; any spectrum change within the
timescale of differential time delays may prevent one from selecting
such systems as lens candidates. We show conditional probability
distributions of time delays proposed by \citet{oguri02a} in Figure
\ref{fig:tdelay}. We find that the dependence of time delays on the
density profile is weak when the separation is large, although this
tendency was already shown by 
\citet{oguri02a}. They also concluded that time delay probability
distribution is insensitive to cosmological parameters. Figure
\ref{fig:tdelay} suggests that lenses with even smaller separations,
$\theta\sim 30''$, are likely to have time delays larger than ten years
which may be typical time scale forming the broad absorption line
\citep{ma02}. Therefore, to assert quasar pairs as lensed images, one
must compare spectral signature which would be unchanged within possible
time delays. Needed time scales are $\Delta t>100{\rm yr}$ for
$\theta\sim 30''$ and $\Delta t>1000{\rm yr}$ for $\theta\sim 200''$. We
note that the information of flux ratio and the central core image may
become one of the evidence of gravitational lensing \citep{rusin02}. 

%%%%%%%%%%%%%%%%%%%%%%%%%%%%%%%%%%%%%%%%%%%%%%%%%%%%%%%%%%
%%%%%%%%%%%%%%%%%%%%%%%%%%%%%%%%%%%%%%%%%%%%%%%%%%%%%%%%%%%
\section{Summary}
%%%%%%%%%%%%%%%%%%%%%%%%%%%%%%%%%%%%%%%%%%%%%%%%%%%%%%%%%%%
%%%%%%%%%%%%%%%%%%%%%%%%%%%%%%%%%%%%%%%%%%%%%%%%%%%%%%%%%%%
In this paper, we have calculated the predicted numbers of
arcminute-separation ($\theta>30''$) lensed quasars in the 2dF QSO
survey. We have presented realistic predictions based on the CDM scenario
taking account of the redshift and magnitude distributions of the quasar
catalog. Detailed comparison between theoretical and observed numbers of 
lensed quasars indicates that the detection of wide separation lenses
puts interesting constraints on the density profile and abundance of dark
halos. The case that only one of six pairs is genuine lens system is
marginally consistent with the model that has cuspy inner density
profile $\rho\propto r^{-1.5}$ and the large value of $\sigma_8$,
$\sigma_8\ga 1.2$ for $\Omega_0=0.3$. To reconcile this result with
X-ray or shear measurement, much smaller $\Omega_0$ ($\Omega_0\la0.15$)
and much larger $\sigma_8$ ($\sigma_8\ga1.5$) are needed. Our result of
this large $\sigma_8$ is similar to that of the Sunyaev-Zel'dovich
angular power spectrum, $\sigma_8\sim 1.1$ \citep{komatsu02}. We have
found also that it is quite hard to produce lenses with separation
$>200''$. Thus a conservative interpretation of this observation is that
none of these quasar pairs is lensed. But if it turns out that some of
these quasar pairs are genuine lens systems, we can put interesting
constraints on not only the density profile of dark halos but also the
$\Omega_0$ and $\sigma_8$ that are somewhat different from X-ray/shear
constraints.  We note that the number of genuine lens systems may be
significantly larger than six which \citet{miller03} reported because
only 11 of 38 candidates has been observed spectroscopically. However,
such anomalously high lensing rate cannot be reproduced by even the most
optimistic models. In this case, we have to examine whether we miss
some important systematic effects which increase lensing rates. One
possible systematic effect is the asymmetry of lensing halos,
although this effect has been considered to be small so far. In any
case, statistics of wide separation lensing offer a promising way to
probe the abundance and density profile of dark halos. 

%%%%%%%%%%%%%%%%%%%%%%%%%%%%%%%%%%%%%%%%%%%%%%%%%%%%%%%%%%%%%%
%%%%%%%%%%%%%%%%%%%%%%%%%%%%%%%%%%%%%%%%%%%%%%%%%%%%%%%%%%%%%%
\section*{Acknowledgments}
%%%%%%%%%%%%%%%%%%%%%%%%%%%%%%%%%%%%%%%%%%%%%%%%%%%%%%%%%%%%%%
%%%%%%%%%%%%%%%%%%%%%%%%%%%%%%%%%%%%%%%%%%%%%%%%%%%%%%%%%%%%%%%
The author would like to thank Yasushi Suto and Naohisa Inada for useful
discussions and comments.
%%%%%%%%%%%%%%%%%%%%%%%%%%%%%%%%%%%%%%%%%%%%%%%%%%%%%%%%%%%%%%%%%%%%%%%
%\clearpage
%%%%%%%%%%%%%%%%%%%%%%%%%%%%%%%%%%%%%%%%%%%%%%%%%%%%%%%%%%%%%%%%%%%%%%%

\label{lastpage}

\end{document}